# Experimental and numerical study on the influence of extra-depth on cut blasting post-blast damage

*Changda Zheng*[1], *Renshu Yang*[1], *Chenxi Ding*[1], *Songlin He*[1], *Bo Wang*[2], *and Yongzhong Yuan*[3]

1) School of Future Cities, University of Science and Technology Beijing, Beijing 100083, China

2) School of Mechanics and Civil Engineering, China University of Mining and Technology (Beijing), Beijing 100083, China

3) Laizhou Huijin Mining Investment Co., Ltd., Yantai 261400, Shandong, China

**Abstract:** Cutting is a key factor affecting the speed of blasting excavation. With the continuous advancement of deep-hole blasting technology, determining the optimal extra-depth of the cut relative to the non-cut blast hole is of paramount importance. By combining model experiments and numerical simulations, this study systematically investigates the effect of extra-depth on post-blast damage morphology, damage width, depth, area, and fractal damage. Furthermore, numerical simulations were employed to validate the experimental results from the perspectives of damage evolution and peak pressure at monitoring points. The main findings are as follows: The extra-depth of the cut has a significant nonlinear effect on the blasting outcome. As the extra-depth increases, blasting damage initially increases but decreases beyond a critical depth due to marginal effects, leading to the appearance of residual blast hole features. Under experimental conditions, when the extra-depth was 15 mm, the damage depth, width, and area reached their maximum values of 43.5 mm, 109.9 mm, and 6055.2 mm², respectively, indicating optimal blasting performance. The fractal damage of the specimen also exhibited a significant trend of initially increasing and then decreasing. The maximum fractal damage, obtained by deriving the fitting curve, was 0.75, corresponding to an extra-depth of 13.7 mm. The numerical simulation results are in good agreement with the experimental findings, showing a significant downward shift in the peak pressure points in the damage zone with increasing extra-depth. In summary, an appropriate extra-depth can achieve optimal borehole utilization, while an excessive extra-depth can lead to residual blast hole formation at the bottom of the cut, reducing effectiveness. This study provides theoretical guidance for optimizing the design parameters of extra-depth in deep-hole blasting.

Corresponding author: Changda Zheng
E-mail: zhengcd95@163.com

# 1 Introduction

Blasting technology [1-2] is one of the commonly used rock fragmentation methods in the engineering field, extensively applied in various areas such as mining operations [3], tunnel excavation [4], and hydraulic construction [5]. Due to its efficient rock-breaking capability and low cost, it has become the preferred method for both underground [6] and open-pit [7] blasting projects. In blasting operations, the cut blasting plays a critical role in determining the overall excavation speed [8-9]. The cut blasting creates a secondary free surface and expansion space within the rock mass, facilitating smoother and more efficient subsequent blast hole operations. The proper design of cut blasting parameters is crucial for achieving the desired blasting outcome, as it not only significantly enhances the excavation speed but also effectively reduces explosive consumption and construction costs.

Factors influencing the effectiveness of cut blasting include, but are not limited to, the free face and expansion space, physical and mechanical properties of the rock, cut design parameters, in-situ stress, and the type and amount of explosives used. Numerous studies have been conducted by blasting scholars on these factors. Regarding free face and expansion space, Zheng *et al.* [10] conducted experimental and numerical investigations on blasting damage characteristics of a medium with a single free boundary and obtained a quantitative relationship between the burden and blasting energy utilization efficiency. Wang *et al.* [11] employed a coupled numerical approach combining LS-DYNA (a transient dynamic finite element program) and UDEC (an universal discrete element code) to compare the dynamic fracture processes in two types of jointed rock masses, examining the influence of free face on crack propagation. Chen *et al.* [12] explored the application of double wedge cuts in improving cut blasting efficiency for medium-depth blasting, analyzing the cavity formation mechanisms of single and double wedge cuts. Their findings showed that the stress wave from the first wedge cut causes pre-damage to the rock mass in the cavity of the second wedge cut, enhancing rock fragmentation and contributing to the formation of a new free surface. Regarding the physical and mechanical properties of rocks, Bhatawdekar *et al.* [13] systematically collected rock mass characteristic data and used artificial neural networks to predict key performance indicators for drilling and blasting, ultimately improving drilling and blasting efficiency. Haghnejad *et al.* [14] utilized a 3D discrete element code to simulate the effects of blast-induced seismic waves on mine slope stability, examining the comprehensive impact of rock physical and mechanical properties on blasting damage. Concerning cut design, Li *et al.* [15] studied the influence of cut blasting parameters, such as hole spacing and cavity diameter, on cut efficiency in a one-step shaft sinking. The results showed that optimizing hole spacing improves cut efficiency, while increasing cavity diameter further enhances cutting performance. Qiu *et al.* [16] investigated short-delay blasting with a single free surface in underground mines and found that it improved rock fragmentation and controlled blast-induced vibrations. Regarding in-situ stress, Yan *et al.* [17] conducted experimental research on crater formation of sandstone specimens under different biaxial compressive stresses to investigate the role of stress on crack propagation, crater characteristics, and fragment distribution. The study found that biaxial static stress suppresses the formation of radial cracks and promotes the development of circumferential cracks, increasing crater diameter, area, and volume, and altering the initial crack morphology. Ding *et al.* [18] utilized theoretical calculations and numerical simulations to investigate the propagation characteristics of blast-induced cracks under different in-situ stress conditions and proposed an improved CDEM method suitable for studying crack propagation in blasting. The findings showed that in-situ stress conditions have a significant impact on the distribution of blast-induced cracks and stress evolution, with biaxial unequal in-situ stress having varying inhibitory effects on horizontal and vertical crack propagation. Regarding the type and amount of explosives, Wei *et al.* [19] simulated the effects of explosive properties, rock mass quality, and loading rate on rock damage. Zhang *et al.* [20] experimentally studied the blasting fragmentation of granite cylinders with different charge amounts. The results indicated that increasing the charge amount improves fragmentation effectiveness, regardless of stemming use.

Although significant progress has been made in the study of cut blasting, some factors have not received sufficient attention. For instance, the concept of "extra-depth," which refers to the depth difference between the cut hole and the non-cut blast hole, is typically considered to be greater than or equal to zero. Extra-depth is also one of the key parameters influencing blasting performance. In practical engineering applications, the extra-depth is often designed as 200 mm, and this design has long been a consensus among blasting engineers. However, with the advancement of mine mechanization and blasting technology, the single-round advance in blasting excavation has significantly increased. For instance, in shaft sinking, the average single-round advance has increased from 2.93 meters in the 1970s to 4.31 meters today, and even reaches 5 to 6 meters in some cases. The increase in single-round advance raises the question of whether it remains reasonable to continue designing the extra-depth as a fixed value, which warrants further scientific investigation. Currently, there is still a lack of research and discussion in this area.

To investigate the influence of extra-depth on the performance of cut blasting, this study employs model experiments and numerical simulations, using fractal theory as the damage evaluation method. The study systematically examines the effect of extra-depth on post-blast fracture morphology, damage width, depth, area, and fractal damage. Numerical simulations are further used to verify the experimental results from the perspectives of damage evolution and peak pressure. The findings provide a scientific basis for optimizing the design parameters of extra-depth in deep hole blasting.

## 2 Experimental study on cut blasting with different extra-depth

## 2.1 Experimental design

Fig. 1 shows the schematic diagram of the specimen and the explosive. Plexiglass was selected as the material for the deep slot model experiments, and the specimen's geometric dimensions were 300 mm × 300 mm × 3 mm (length × width × thickness). A strip-shaped blasthole with a diameter of 3 mm was prefabricated along the midline of the top edge of the specimen as the cut hole, with a length of $H_c$. Another strip-shaped blasthole with the same diameter was prefabricated 17 mm to the left of the cut hole as the non-cut blast hole, with a length of $H_n$=30 mm. The difference between $H_c$——the length of the cut hole and $H_n$——the length of the non-cut blast hole is defined as $H_e$——the extra-depth. The loading parameters for the cut and non-cut blast holes were identical; lead azide was used as the explosive, with a mass of 100 mg and a length of 21 mm. The initiation method was reverse initiation, and the delay time between the cut and non-cut blast holes was set to 15 ms.

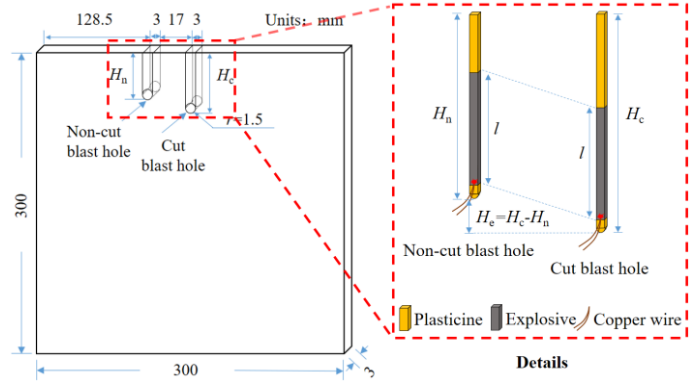

Fig. 1. The schematic diagram of the specimen and explosive

Twenty-one experimental groups were designed using extra-depth $H_e$ as the sole variable, ranging from 0 mm to 30 mm with increments of 1.5 mm. Each experimental setup was repeated twice to ensure reproducibility of the results. During the experiment, plasticine was used to secure the explosive within the borehole, and a detonator probe was inserted into the explosive with a lead wire connected to the blasting device for initiation.

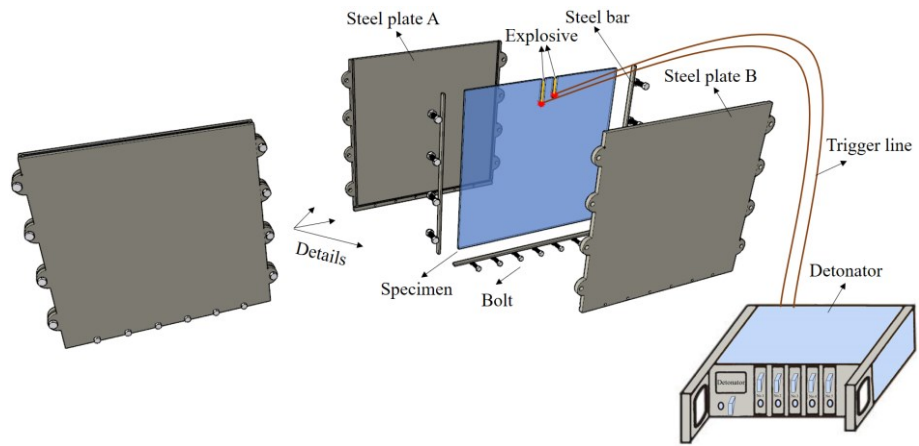

Fig. 2. Schematic diagram of the blasting model experiment system

Boundary conditions significantly influence the generation and propagation of blast-induced cracks. To approximate the single-free-surface condition of rock mass cutting in engineering blasting, the experimental apparatus depicted in Fig. 2 was designed to constrain the specimen. All components of the setup were fabricated using 405 stainless steel to meet the requirements of structural rigidity and corrosion resistance during the experiment. During operation, the specimen was first clamped between two steel plates, followed by inserting steel bars into the grooves along the steel plate edges. Finally, the steel plates were fixed using bolts. A digital torque wrench was employed to ensure that all bolts had consistent torque, thereby achieving uniform stress distribution within the specimen. This setup ensured that the boundary conditions of the specimen closely matched those in actual engineering blasting, which is crucial for the successful execution of the experiment.

## 2.2 Analysis of post-blast fracture morphology

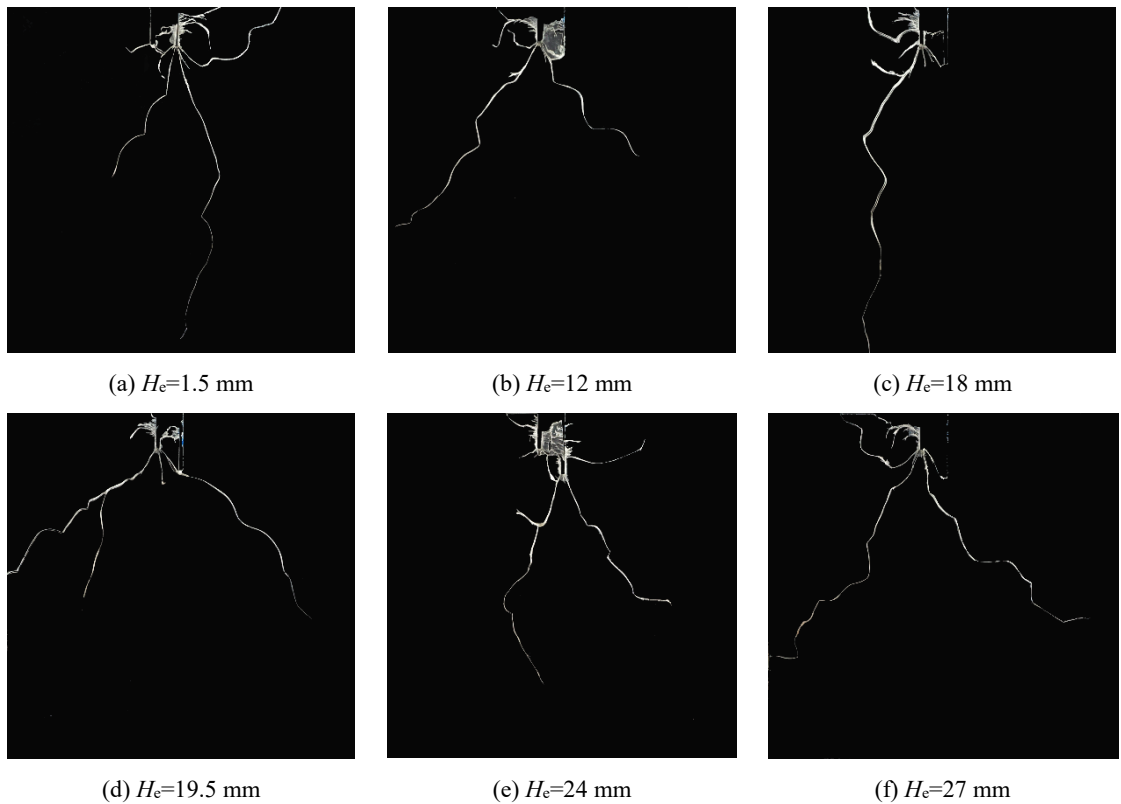

(a) $H_e$=1.5 mm    (b) $H_e$=12 mm    (c) $H_e$=18 mm

(d) $H_e$=19.5 mm    (e) $H_e$=24 mm    (f) $H_e$=27 mm

Fig. 3. Invalid experimental results

By comparing the post-blast damage morphology of each group of specimens, it was found that six experimental results in Fig. 3 were invalid, as no damage occurred on the right side of the cutting hole or the left side of the non-cut blast hole. These results were identified as outliers and subsequently excluded. The cause may be improper handling during explosive wiring, resulting in low charge density or poor contact of the lead wire, leading to a failure in detonation.

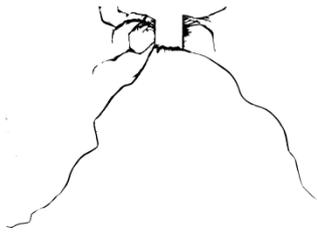
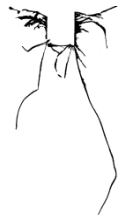
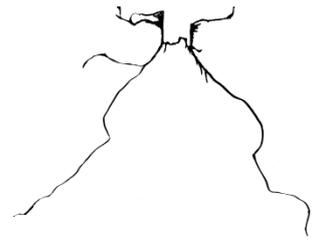

(a) $H_e$=0 mm  (b) $H_e$=3 mm  (c) $H_e$=4.5 mm

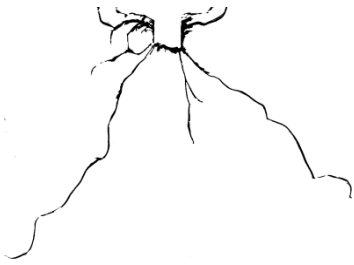
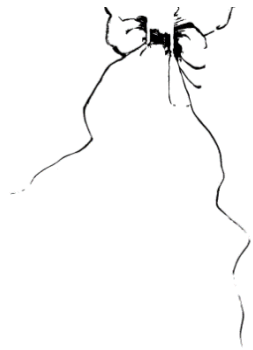
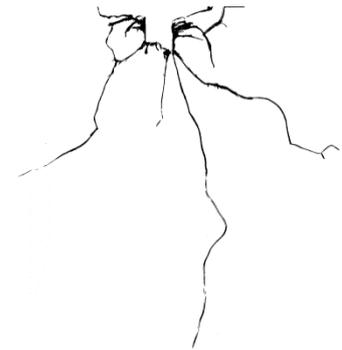

(d) $H_e$=6 mm  (e) $H_e$=7.5 mm  (f) $H_e$=9 mm

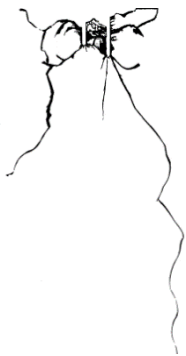
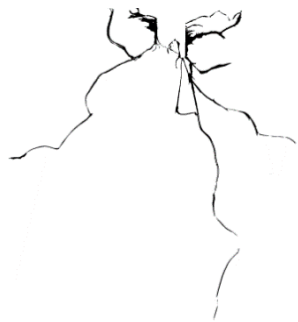
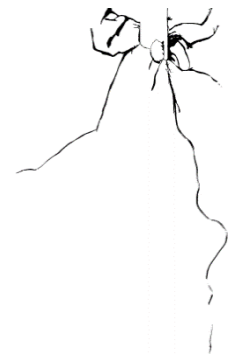

(g) $H_e$=10.5 mm  (h) $H_e$=13.5 mm  (i) $H_e$=15 mm

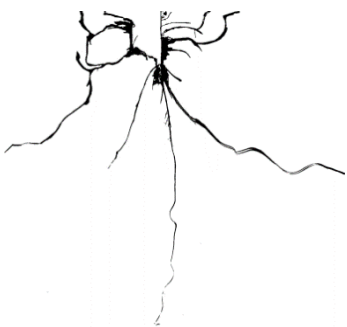
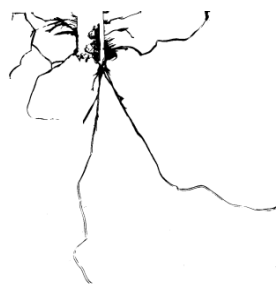
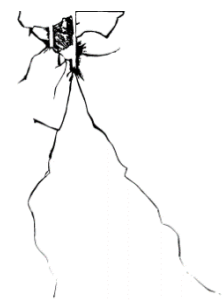

(j) $H_e$=16.5 mm  (k) $H_e$=21 mm  (l) $H_e$=22.5 mm

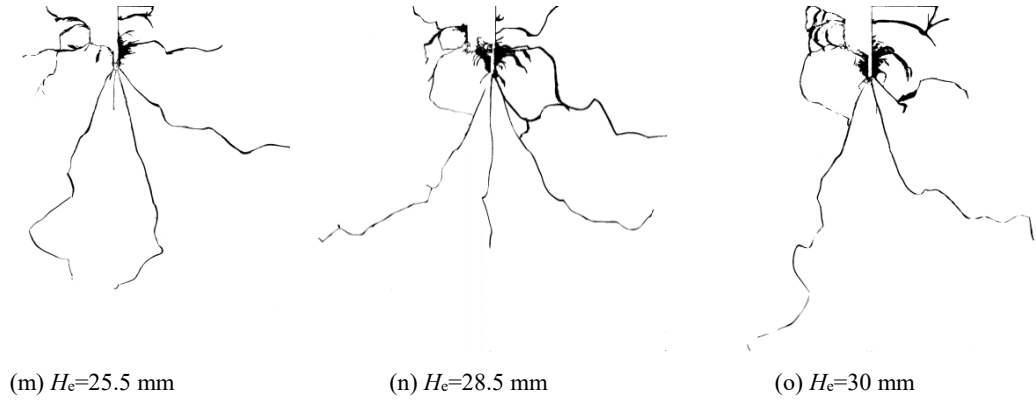

  (m) $H_e$=25.5 mm      (n) $H_e$=28.5 mm      (o) $H_e$=30 mm

Fig. 4. Binary images of specimen after explosion

  After excluding the six invalid data sets, the remaining 15 experimental results were binarized for subsequent evaluation of their blast damage, as shown in Fig. 4. The binarized images have dimensions of 1500×1500 pixels, with a resolution of 138 pixels/cm. In Fig. 4, the cutting holes and non-cut blast holes of specimens with different extra-depths were successfully detonated, resulting in damage on both sides. Blast-induced cracks propagated from the borehole walls towards the upper free surface, creating fragments of various sizes, some of which were ejected from the parent material. Extensive damage occurred between the cutting and non-cut blast holes due to the explosion, with a large number of fragments being ejected and effectively unrecoverable. Additionally, 2 to 4 long blast-induced cracks were generated at the bottom of the cutting and non-cut blast holes, attributed to stress concentration at the bottom of the elongated borehole and the structural effects caused by the small size of the specimen. Furthermore, at smaller extra-depths, the bottoms of the cutting and non-cut blast holes tended to connect as shown in Fig. 4(a)~(i)), leading to near-complete destruction of the entire borehole. As the extra-depth increased, residual blasthole features began to appear, as shown in Fig. 4(j)~(o).

## 2.3 Analysis of post-blast damage width, damage depth, and damage area

  In this experiment, in addition to the crushed zone and the fracture zone, the blast-induced cracks, influenced by the free surface, fully extended and interacted with the free surface to form multiple fragments, with many fragments detaching from the specimen's main body. This region, where fragments are detached from the main body, is defined as the ejection zone. The crushed zone, fracture zone, and ejection zone collectively form the damage region of the cutting blast. For convenient quantitative comparison, the distance from the lower boundary of the damaged region to the upper boundary free surface is defined as the damage depth, measured in mm. The distance from the left boundary to the right boundary of the damaged region is defined as the damage width, also in mm. The area of the damaged region is defined as the damage area, measured in mm². MATLAB was used to calculate the damage width, depth, and area of the specimen after blasting.

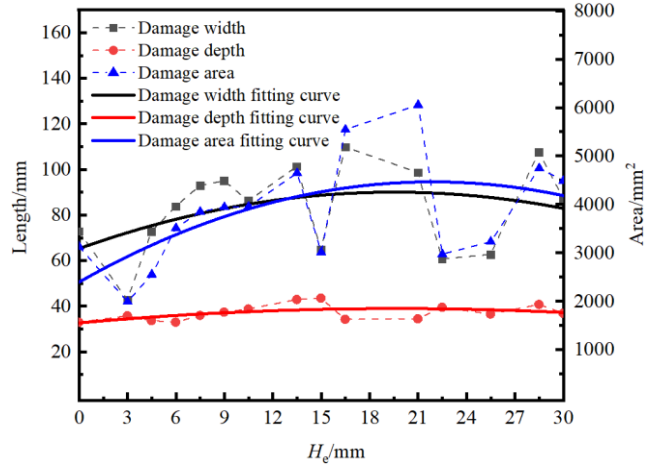

Fig. 5. Curves of blasting damage width, damage depth, and damage area versus $H_e$ for specimen

Fig. 5 presents the scatter plots and quadratic polynomial fitting curves for damage width, damage depth, and damage area versus extra-depth after blasting. Analysis of the scatter plots indicates that the damage depth reaches its maximum value of 43.5 mm when $H_e$ = 15.0 mm, the damage width reaches its maximum value of 109.9 mm when $H_e$=16.5 mm, and the damage area reaches its maximum value of 6055.2 mm² when $H_e$=21.0 mm. From the fitted curves, it can be observed that, with increasing extra-depth, the damage depth, width, and area all show a trend of first increasing and then decreasing. By deriving the quadratic polynomial fitting curves for damage depth, width, and area to find the peaks, it is found that the damage depth reaches a maximum value of 38.6 mm, the damage width reaches a maximum value of 88.7 mm, and the damage area reaches a maximum value of 4262.0 mm², all at $H_e$=15.0 mm. The quantitative analysis results are consistent with the previous analysis of the specimen's failure modes. When $H_e$ < 15.0 mm, the damage depth, width, and area all increase with increasing $H_e$. When $H_e$ > 15.0 mm, residual blastholes begin to appear, and the damage depth, width, and area decrease accordingly. Overall, considering the damage width, depth, and area of the blasted specimen, an extra-depth of $H_e$ = 15.0 mm is determined to be optimal for this experiment, corresponding to an extra-depth coefficient of 0.5.

## 2.4 Analysis of post-blast fractal damage

Quantitative characterization methods of rock damage include CT scanning [21], acoustic emission [22], ultrasonic testing [23], and fractal theory [24]. Among them, one approach involves using the fractal dimension to represent the extent of damage. By combining fractal theory with damage characterization, the concept of fractal damage (denoted as $\omega$) is introduced. This method is characterized by its low cost, high speed, and simplicity of operation, and it is chosen in this study to evaluate the damage of the specimens after blasting. Under explosive loading, the relationship between fractal damage $\omega$ and the fractal dimension $D$ can be expressed as follows:

$$\omega = \frac{D_t - D_0}{D_t^{max} - D_0} \tag{1}$$

In the equation: $\omega$ represents the fractal damage under explosive loading; $D_t$ is the fractal dimension of the damaged region after blasting; $D_0$ is the fractal dimension of the damaged region before blasting; $D_t^{max}$ is the fractal dimension in the case of complete damage, where $D_t^{max} = 2$ for two-dimensional problems, and $D_t^{max} = 3$ for three-dimensional problems.

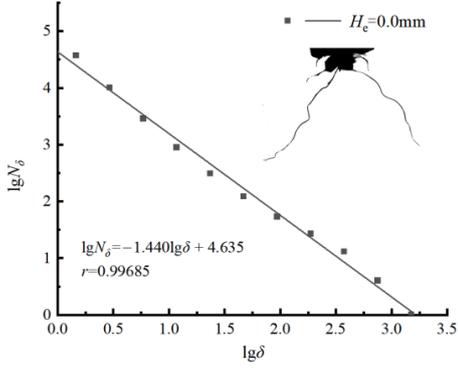

(a) $H_e$=0 mm

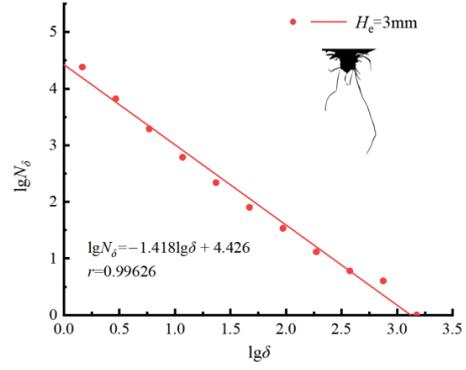

(b) $H_e$=3 mm

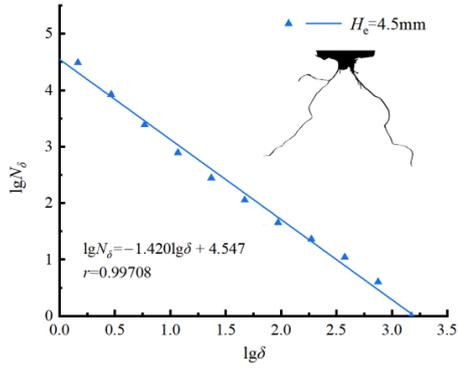

(c) $H_e$=4.5 mm

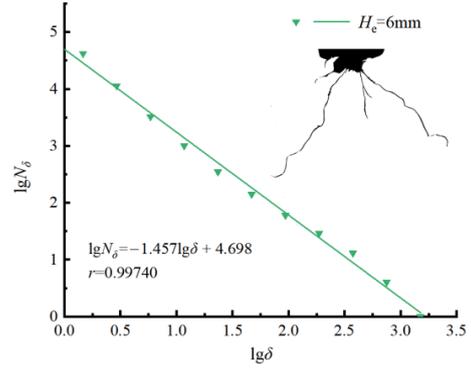

(d) $H_e$=6 mm

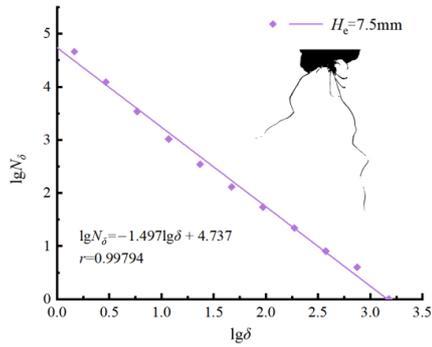

(e) $H_e$=7.5 mm

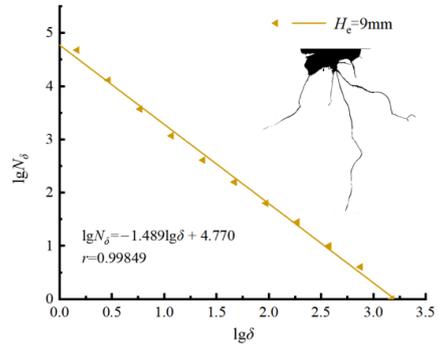

(f) $H_e$=9 mm

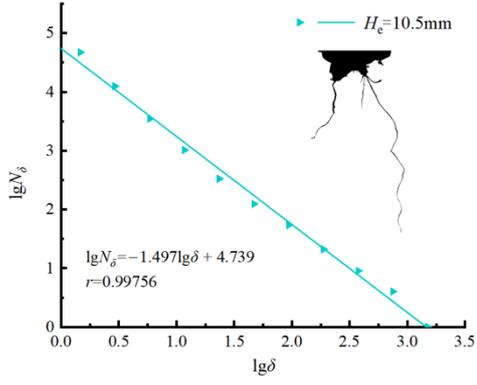

(g) $H_e$=10.5 mm

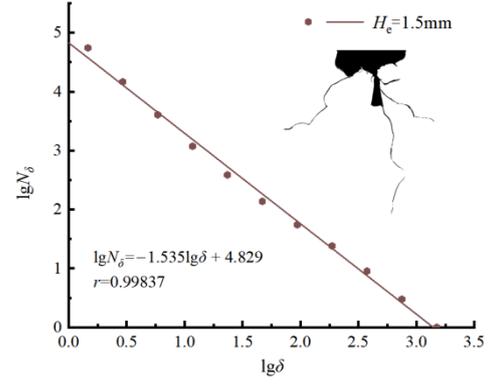

(h) $H_e$=13.5 mm

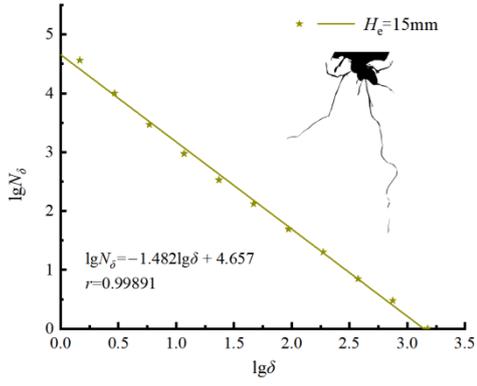

(i) $H_e$=15 mm

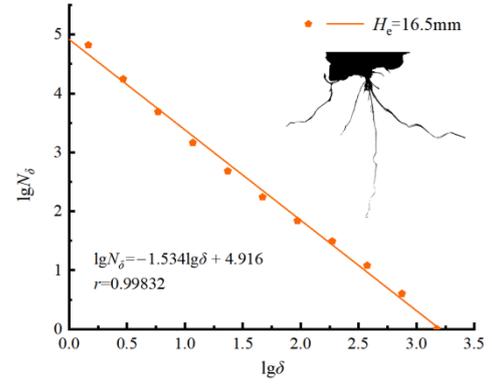

(j) $H_e$=16.5 mm

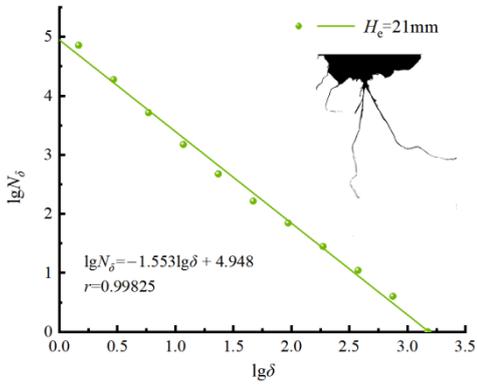

(k) $H_e$=21 mm

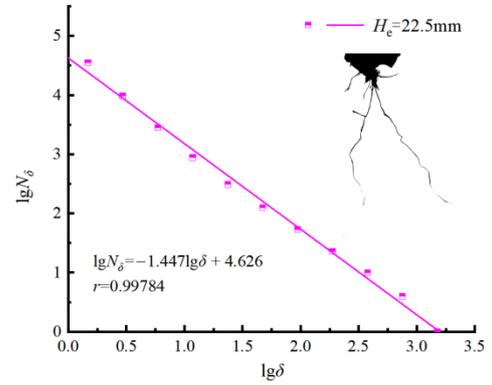

(l) $H_e$=22.5 mm

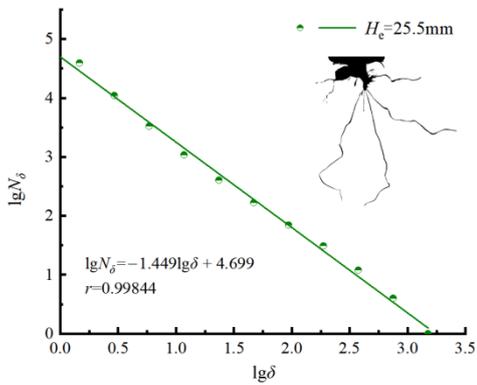

(m) $H_e$=25.5 mm

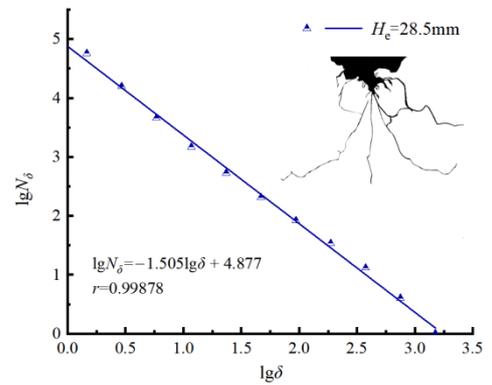

(n) $H_e$=28.5 mm

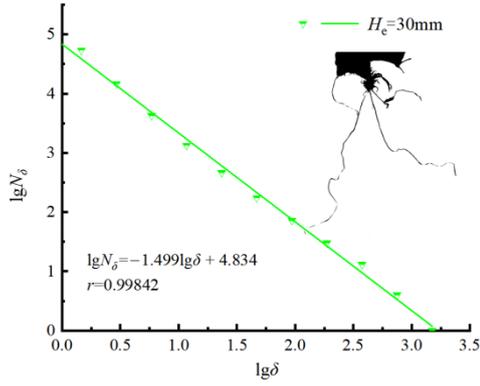

(o) $H_e$=30 mm

Fig. 6. Fitting lines of box-counting dimension for blasting damage zone of specimen

The crushed zone, fracture zone, and ejection zone in the post-blast images were all painted black and then imported into the fractal analysis software for processing. The fitting lines for the box-counting dimension of the damage regions in the post-blast specimens were obtained, as shown in Fig. 6 The correlation coefficients of the fitting lines were all greater than 0.99, indicating a high degree of data fitting. The slopes of the fitting lines were recorded to determine the box-counting dimensions of the post-blast specimens for different extra-depth, as shown in Table 1.

Table 1. Fractal dimension of the specimen after blasting

| $H_e$/mm | 0 | 3 | 4.5 | 6 | 7.5 | 9 | 10.5 | 13.5 |
|---|---|---|---|---|---|---|---|---|
| Fractal dimension | 1.440 | 1.418 | 1.420 | 1.457 | 1.497 | 1.489 | 1.497 | 1.535 |
| $H_e$/mm | 15 | 16.5 | 21 | 22.5 | 25.5 | 28.5 | 30 | |
| Fractal dimension | 1.482 | 1.534 | 1.553 | 1.446 | 1.449 | 1.505 | 1.499 | |

The specimen was uniform and undamaged before blasting, thus $D_0$=0. Based on Equation 1 and Table 1, the scatter plot and cubic polynomial fitting curve of fractal blasting damage $\omega$ versus extra-depth $H_e$ are obtained, as shown in Fig. 7.

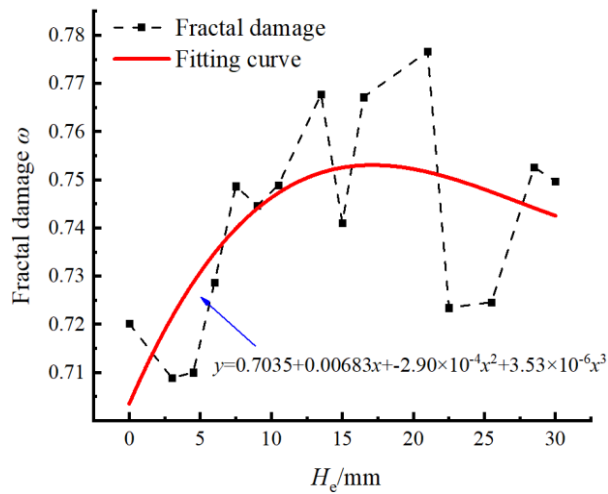

Fig. 7. Relationship curve between fractal blasting damage $\omega$ and $H_e$

From Fig. 7, it can be observed that when $H_e$=3 mm, the fractal blasting damage reaches its minimum value 0.709, whereas at $H_e$=21 mm, the fractal blasting damage attains its maximum value 0.7765. The cubic polynomial fitting curve of the fractal blasting damage suggests that, as the extra-depth increases, the blasting damage first increases and then decreases. By deriving the curve to find the peak, it is determined that the fractal blasting damage reaches its maximum value 0.75 at $H_e$=13.7 mm, corresponding to an extra-depth factor of 0.46.

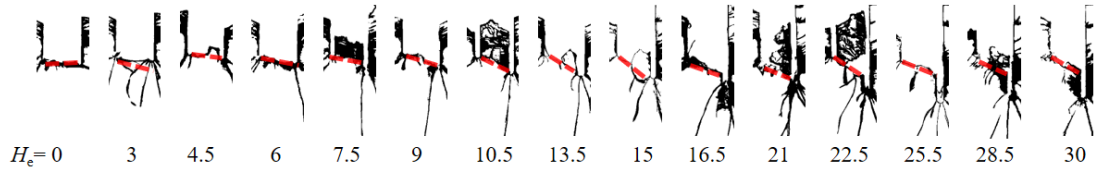

Fig. 8. Connecting line at the bottom of the cavity between the cutting hole and the non-cut blast hole after blasting

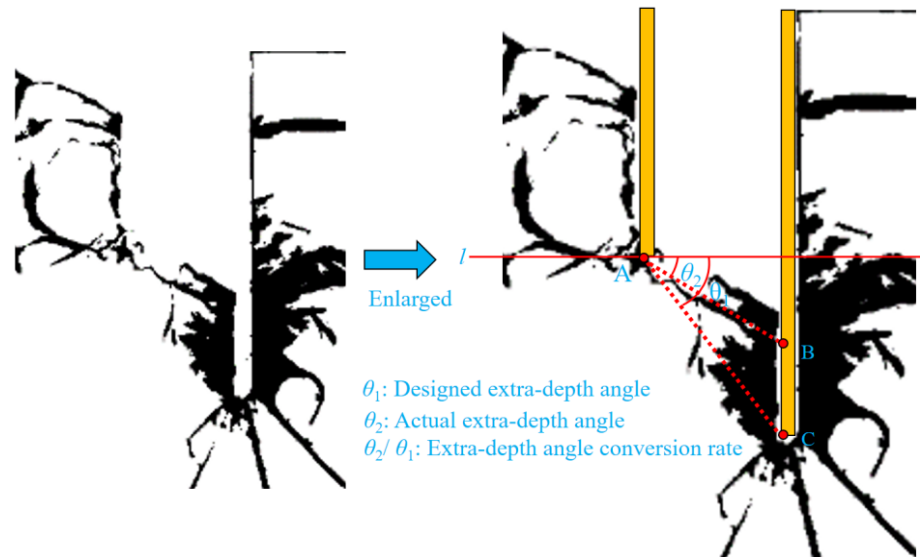

Fig. 9. Definition of designed extra-depth angle, actual extra-depth angle, and extra-depth angle conversion rate (taking $H_e$=30 mm as an example)

Fig. 8 shows the connection between the bottom of the cutting hole and the non-cut blast hole cavity for specimens with different extra-depth after blasting. To further quantify the appearance of residual blastholes with increasing extra-depth, the post-blasting result at $H_e$=30 mm is taken as an example, as shown in Fig. 9. The concepts of designed extra-depth angle, actual extra-depth angle, and extra-depth angle conversion rate are defined. The designed extra-depth angle $\theta_1$ is defined as the acute angle between line AC, connecting point A at the bottom of the non-cut blast hole and point C at the bottom of the cutting hole, and horizontal line L. The actual extra-depth angle $\theta_2$ is defined as the acute angle between line AB, connecting point A at the bottom of the non-cut blast hole and point B at the upper end of the residual blast hole in the cutting hole, and horizontal line L. The extra-depth angle conversion rate is defined as the ratio of the actual extra-depth angle $\theta_2$ to the designed extra-depth angle $\theta_1$.

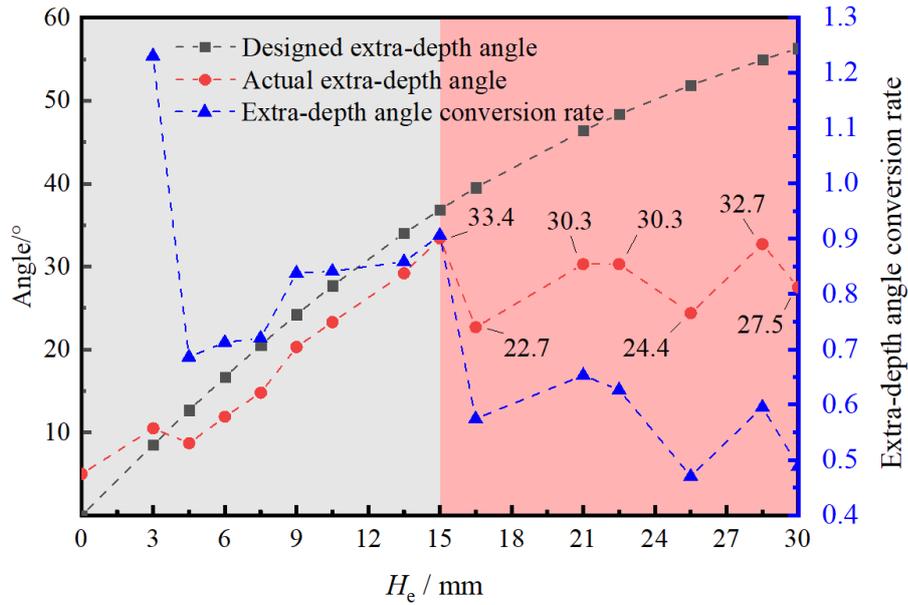

Fig. 10. Scatter plot of designed extra-depth angle, actual extra-depth angle, and extra-depth angle conversion rate *vs* $H_e$

Fig. 10 presents the scatter plot of designed extra-depth angle, actual extra-depth angle, and extra-depth angle conversion rate versus He. As shown in the figure, with the increase in $H_e$, the designed extra-depth angle $\theta_1$ continuously increases. When $H_e$<15 mm, the actual extra-depth angle gradually increases with He. When $H_e$>15 mm, the actual extra-depth angle fluctuates around 28.8° (the average value of the actual extra-depth angle for the last 7 groups when $H_e$>15 mm). When $H_e$=0 mm, the designed extra-depth angle is 0°, and according to the definition, the extra-depth angle conversion rate does not exist. When $H_e$=3 mm, due to the presence of the crushed zone, the extra-depth angle conversion rate is greater than 1. When 333 mm < $H_e$<15 mm, the conversion rate is less than 1, and it fluctuates upwards with the increase in He . When $H_e$>15 mm, the conversion rate remains less than 1 and fluctuates downwards as He increases.

## 3 Numerical simulation of cut blasting with different extra-depth
### 3.1 Establishment of numerical model and design of case

Using the LS-DYNA software, five numerical models were established, named Case-1 to Case-5. The dimensions of the numerical models were set to be the same as those of the experimental models, i.e., 300 mm × 300 mm, with a thickness of one element, resulting in a total of 178,864 tetrahedral elements. The keyword selection and parameter settings for the explosive and rock mass were based on the settings in reference [10]. The top surface of the model was left free to simulate the free surface condition, while the left, right, and bottom surfaces were set as non-reflecting boundaries. The front and back surfaces were fixed. The explosive and rock mass were coupled through fluid-solid interaction, where the explosive and air were treated as fluids, and the rock mass was treated as a solid.

Table 2. Design parameters for Case-1 to Case-5

|        | Case-1 | Case-2 | Case-3 | Case-4 | Case-5 |
|--------|--------|--------|--------|--------|--------|
| $H_b$/mm | 21 | 21 | 21 | 21 | 21 |
| $H_e$/mm | 0  | 6  | 15 | 24 | 30 |
| $H_n$/mm | 30 | 30 | 30 | 30 | 30 |
| $H_c$/mm | 30 | 36 | 45 | 54 | 60 |

Table 2 provides the design parameters for Case-1 to Case-5, where $H_b$ represents the explosive length, $H_n$ represents the length of non-cut blast hole, and $H_c$ represents the length of cut blast hole. The initiation method was set to be inverse initiation, with the initiation point being the geometric center of the lowest element in the explosive section. The cutting hole was detonated first, followed by the non-cut blast hole, with a delay time of 15 ms. Taking Case-1 as an example, monitoring point A was positioned at the midpoint of the line connecting the bottoms of the cutting hole and the non-cut blast hole. Points B to F were arranged vertically downward from point A at 7.5 mm intervals. The monitoring point arrangement for the other cases was identical to that for Case-1, as shown in Fig. 11.

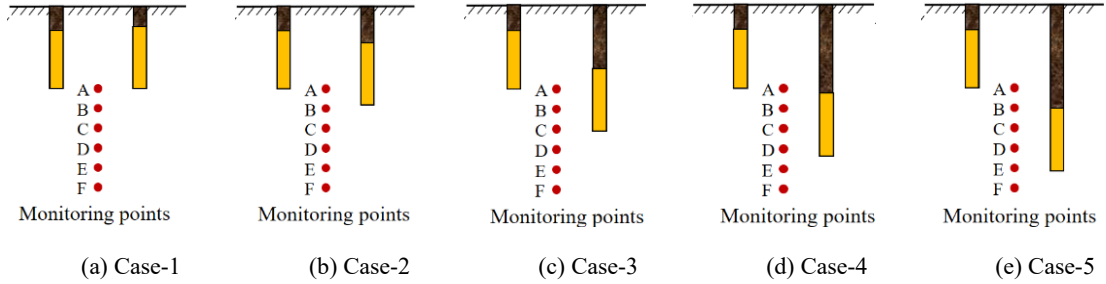

(a) Case-1　　(b) Case-2　　(c) Case-3　　(d) Case-4　　(e) Case-5

Fig. 11. Schematic diagram of monitoring point arrangement for Case-1 to Case-5

### 3.2 Analysis of blasting damage evolution

The numerical simulation results of blasting damage for Case-1 to Case-5 are shown in Fig. 12. It can be observed from the figure that the numerical simulation results of blasting damage for Case-1 to Case-5 are in good agreement with the experimental results. Overall, post-blast damage is mainly distributed on both sides of the cutting hole and non-cut blast hole, initiating from the bottom of the blast hole and propagating upward toward the free surface. In all cases, severe damage occurs between the cutting and non-cut blast holes, similar to the experimental findings. Furthermore, the linkage phenomenon between the non-cut blast hole and the cutting hole is observed. When He = 0 mm, the damaged region at the bottom of the non-cut blast hole connects to the bottom of the cutting hole. As He increases, the damage region from the non-cut blast hole extends diagonally downward, attracted by the free surface formed by the cutting hole detonating first.

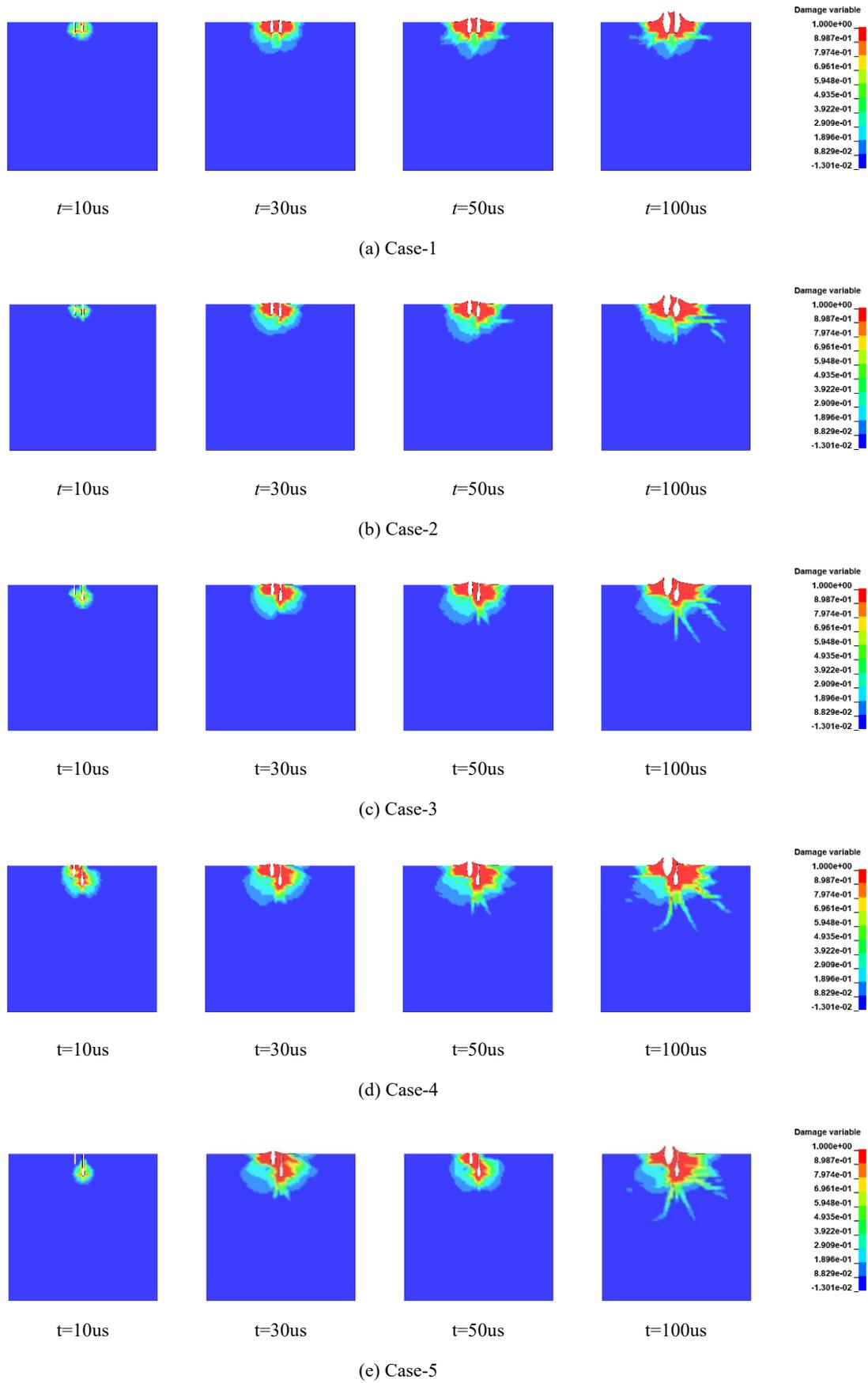

Fig. 12. Evolution process of blasting damage in numerical simulations for Case-1 to Case-5

In Case-1, at t = 30 μs, the bottom of the cutting hole and the non-cut blast hole are interconnected, and the area between the cutting hole and non-cut blast hole is completely destroyed. At t = 50 μs, influenced by the explosive, the area between the cutting and non-cut blast holes starts to eject upward. Both the right side of the cutting hole and the left side of the non-cut blast hole near the free surface also begin to eject upward. The overall damage area resembles the superposition of two funnels, resulting in a symmetric damage distribution. At t = 100 μs, the area of the damage region shows a slower increase compared to t = 50 μs, with the ejection process still ongoing. In Case-2, at t = 30 μs, the damaged area at the bottom of the non-cut blast hole exceeds the depth of the non-cut blast hole, extending diagonally from the bottom of the non-cut blast hole toward the cutting hole. At t = 50 μs, the damage region continues to expand horizontally, with the downward growth of damage slowing, and ejection begins. At t = 50 μs, the overall damage region roughly presents a state of two superimposed funnels—one larger and one smaller—resulting in an asymmetric post-blast damage distribution. Considering the limited space, the damage evolution processes for Case-3 to Case-5 are not detailed individually in this paper.

### 3.3 Analysis of pressure variation curves at monitoring points over time

For Case-1 to Case-5, the pressure variation curves for monitoring points A to F are shown in Fig. 13. As depicted, when $H_e$ = 0 mm (Fig. 13(a)), the peak pressure at point A is the highest, and the peak pressure gradually decreases as the distance from the borehole increases, following the order: A > B > C > D > E > F. When $H_e$ = 6 mm (Fig.13(b)), point A still shows the highest peak pressure. The peak pressure at the six monitoring points follows the order: A > C > B > D > E > F. Compared to $H_e$ = 0 mm, the peak pressure at point C surpasses that of point B above it. When $H_e$ = 15 mm (Fig. 13(c)), point A again exhibits the highest peak pressure. The order of peak pressure at the monitoring points is: A > B > C > D > E > F. When $H_e$ = 24 mm (Fig. 13(d)), point B shows the highest peak pressure at $4.76e^{-3}$, with point C having a pressure value close to that of point B at $4.73e^{-3}$. The order of peak pressures at the monitoring points is: B > C > A > D > E > F. When $H_e$ = 30 mm (Fig. 13(e)), point C has the highest peak pressure, and the order of peak pressure at the six monitoring points is: C > D > B > E > A > F.

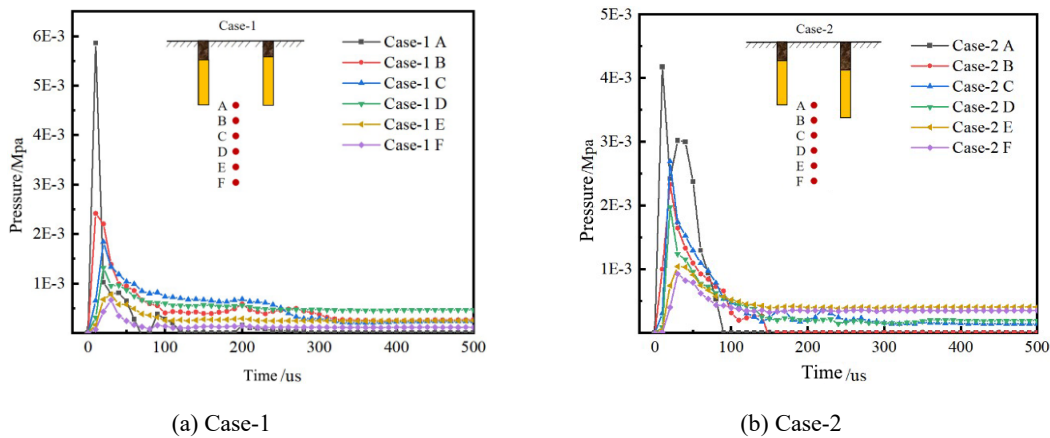

(a) Case-1　　　　　　　　　　　　　(b) Case-2

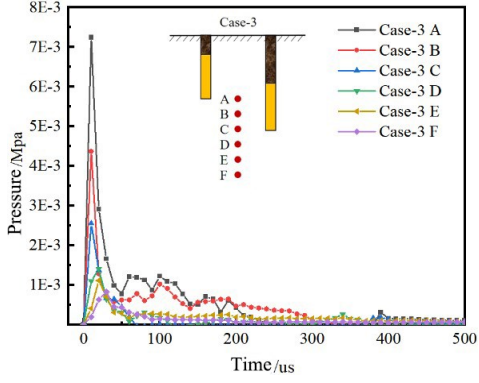
(c) Case-3

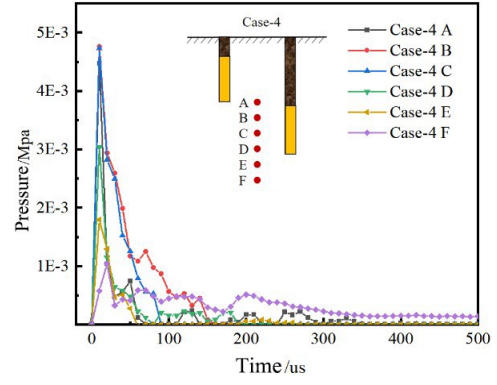
(d) Case-4

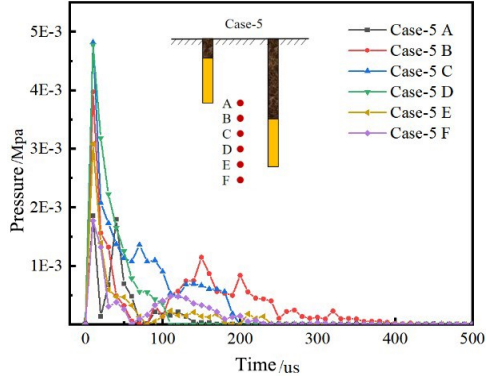
(e) Case-5

Fig. 13. Pressure variation curves for monitoring points A to F in Case-1 to Case-5

Overall, in Case-1 to Case-5, as $H_e$ increases, the locations of the pressure peak measurement points sequentially shift from Point A, Point A, Point A, Point B, to Point C. The increase in the cut depth leads to a downward shift in the location of the pressure peak between the non-cut blast holes and the cut holes. When the pressure exerted on a specific unit exceeds its ultimate pressure capacity, the unit fails, which macroscopically manifests as structural damage. In other words, the increase in the cut hole depth causes the free surface formed by the cut hole to move downward. This downward movement of the free surface, formed first by the cut hole, influences the damage region of the non-cut blast holes to also shift downward at an angle, thereby improving the overall utilization of the blast holes.

## 4 Conclusions

This study aims to explore the effects of different extra-depths in cut blasting on blasting performance, focusing on changes in damage depth, width, area, crack morphology, and fractal damage. Using a 2D plexiglass model experiment and numerical simulations with LS-DYNA software, the mechanisms of rock fragmentation by deep cut blasting, post-blast fractal damage characteristics, and damage evolution patterns were systematically studied. The core objective of this research is to optimize the extra-depth design parameters of cut holes to improve blasting efficiency and provide a scientific basis for engineering applications. The main findings of the

study are as follows:

(1) Nonlinear influence of extra-depth on blasting damage. With increasing extra-depth of the cut hole, the depth, width, and damage area of blasting damage showed a pattern of first increasing and then decreasing. Specifically, at an extra-depth of 15.0 mm, the damage depth, width, and area reached their maximum values, being 43.5 mm, 109.9 mm, and 6055.2 mm², respectively. Further analysis using fitted curves showed that the optimal extra-depth is 15.0 mm, at which point the rock-breaking effect of the cut hole is most significant, achieving the best overall blasting efficiency.

(2) Evolution of damage morphology and fractal characteristics. The experimental results revealed that the fractal dimension of the post-blast specimens showed a distinct trend of first increasing and then decreasing with different extra-depths. The maximum fractal damage of 0.75, obtained through curve fitting and differentiation, corresponds to an extra-depth of 13.7 mm. The fractal analysis results indicate that the rock-breaking capacity of the cut hole gradually improves with increasing extra-depth, but beyond the critical depth, the complexity of the damage begins to decline, leading to the occurrence of residual blast hole phenomena, which adversely affects the rock-breaking performance of the cut hole.

(3) Consistency between numerical simulation and experimental results. The numerical simulation results agree well with the experimental results, revealing the evolution process of blasting damage under extra-depth cutting. With increasing extra-depth, the peak pressure point in the damage area shifts significantly downward. When the extra-depth reaches an appropriate range, optimal utilization of the borehole can be achieved; however, excessive extra-depth results in the appearance of residual blast hole phenomena, which hinders the improvement of the cut performance.

## Acknowledgements

The authors would like to acknowledge the anonymous reviewers for their valuable and constructive comments. This work was financially supported by the National Natural Science Foundation of China (No. 52204085 and No. 51934001)

## Conflict of Interest

The authors confirm that they have no commercial or associative interests that could present a conflict of interest related to the submitted work.

## References

[1] Z.X. Zhang, D.F. Hou, Z.R. Guo, Z.W. He, and Q.B. Zhang, Experimental study of surface constraint effect on rock fragmentation by blasting, *Int. J. Rock Mech. Min. Sci.*, 128(2020), art. No. 104278. https://doi.org/10.1016/j.ijrmms.2020.104278

[2] P. Xu, R.S. Yang, J.J. Z, C.X. D, C. Chen, Y. Guo, S.Z. Fang and Y.F. Zhang, Research progress of the fundamental theory and technology of rock blasting, *Int. J. Miner. Metall. Mater.*, 29(2022),

No. 4, p. 705-716. https://doi.org/10.1007/s12613-022-2464-x

[3] F.Y. Liu, W.Y. Chen, Z.Q. Yang, W.X. Deng, H. L and T.H. Yang, Landslide characteristics and stability control of bedding rock slope: a case study in the Sijiaying Open-Pit Mine, *Mining. Metall. Explor.*, 2024(2024), p. 1-16. https://doi.org/10.1007/s42461-024-01110-2

[4] Z.Y. Sun, D.L Zhang, and Q. Fang, Technologies for large cross-section subsea tunnel construction using drilling and blasting method, *Tunn. Undergr. Sp. Tech.*, 141(2023), art. No. 105161. https://doi.org/10.1016/j.tust.2023.105161

[5] R.K. Singh, C. Sawmliana, and P.Hembram, Time-constrained demolition of a concrete cofferdam using controlled blasting, *Innov. Infrastruct. Solut.*, 6(2021), art. No. 20. https://doi.org/10.1007/s41062-020-00387-8

[6] B. He, D.J. Armaghani, S.H. Lai, X.Z. He, P.G. Asteris, and D.C. Sheng, A deep dive into tunnel blasting studies between 2000 and 2023—A systematic review, *Tunn. Undergr. Sp. Tech.*, 147(2024), art. No. 105727. https://doi.org/10.1016/j.tust.2024.105727

[7] Z.D. Leng, Y. Fan, Q.D. Gao, and Y.G. Hu, Evaluation and optimization of blasting approaches to reducing oversize boulders and toes in open-pit mine, *Int. J. Min. Sci. Technol.*, 30(2020), No. 3, p. 373-380. https://doi.org/10.1016/j.ijmst.2020.03.010

[8] S.J. Qu, X.B. Zheng, L.H. Fan, and Y. Wang, Numerical simulation of parallel hole cut blasting with uncharged holes, *Int. J. Miner. Metall. Mater.*, 15(2008), No. 3, p. 209-214. https://doi.org/10.1016/S1005-8850(08)60040-7

[9] A. Aqazddammou, T. Belem, S. Chlahbi, A. Camile, N. Assioui and A. Khalil, Experimental and numerical study on the influence of cut design, deviation factor, and rock mass quality on the blast pull performance, *Geotech. Geol. Eng.*, 42(2024), p. 6601-6623. https://doi.org/10.1007/s10706-024-02914-1

[10] C.D. Zheng, R.S. Yang, Q. Li, C.X. Ding, C.L. Xiao, Y. Zhao, and J. Zhao, Fractal analysis for the blast-induced damage in rock masses with one free boundary, *Mech. Adv. Mater. Struc.*, 31(2022), No. 6, p. 1214-1228. https://doi.org/10.1080/15376494.2022.2134609

[11] Z.L. Wang and H. Konietzky, Modelling of blast-induced fractures in jointed rock masses, *Eng. Fract. Mech.*, 76 (2009), No. 12, p. 1945-1955. https://doi.org/10.1016/j.engfracmech.2009.05.004

[12] B. Cheng, H.B. Wang, Q. Zong, Y. Xu, M.X. Wang, and Q.Q. Zheng, Study of the double wedge cut technique in medium-depth hole blasting of rock roadways, *Arab. J. Sci. Eng.*, 46(2021), p. 4895-4909. https://doi.org/10.1007/s13369-020-05279-8

[13] R.M. Bhatawdekar, M.T. Edy, and J.A. Danial, Building information model for drilling and blasting for tropically weathered rock, *J. Mines Met. Fuels.*, 67(2019), No. 11, p. 494. https://informaticsjournals.co.in/index.php/jmmf/article/view/31661

[14] A. Haghnejad, K. Ahangari, P. Moarefvand, and K. Goshtasbi, Numerical investigation of the

impact of rock mass properties on propagation of ground vibration, *Nat. Hazards.*, 96(2019), p. 587-606. https://doi.org/10.1007/s11069-018-3559-6

[15] Q.Y. Li, K. Liu, X.B. Li, Z.W. Wang, and L. Weng, Cutting parameter optimization for one-step shaft excavation technique based on parallel cutting method, *T. Nonferr. Metal. Soc.*, 28(2018), No. 7, p. 1413-1423. https://doi.org/10.1016/S1003-6326(18)64780-6

[16] X.Y. Qiu, X.Z. Shi, Y.G. Gou, J. Zhou, H. Chen, and X.F. Huo, Short-delay blasting with single free surface: Results of experimental tests, *Tunn. Undergr. Sp. Tech.*, 74(2018), p. 119-130. https://doi.org/10.1016/j.tust.2018.01.014

[17] G.L. Yan, F.P. Zhang, T.S. Ku, Q.Q. Hao, J.Y. Peng, Experimental study and mechanism analysis on the effects of biaxial in-situ stress on hard rock blasting, *Rock. Mech. Rock. Eng.*, 56(2023), p. 3709-3723. https://doi.org/10.1007/s00603-022-03205-y

[18] C.X. Ding, X.T. Liang, R.S. Yang, Z.X. Zhang, X. Guo, C. Feng, X.G. Zhu, and Q.M. Xie, A study of crack propagation during blasting under high in-situ stress conditions based on an improved CDEM method, *Mech. Adv. Mater. Struc.*, 31(2023), No. 20, p. 4922-4939. https://doi.org/10.1080/15376494.2023.2208112

[19] X.Y. Wei, Z.Y. Zhao, and J. Gu, Numerical simulations of rock mass damage induced by underground explosion, *Int. J. Rock Mech. Min. Sci.*, 46(2009), No. 7, p. 1206-1213. https://doi.org/10.1016/j.ijrmms.2009.02.007

[20] Z.X. Zhang, D.F. Hou, Z. Guo, and Z.W. He, Laboratory experiment of stemming impact on rock fragmentation by a high explosive, *Tunn. Undergr. Sp. Tech.*, 97(2020), art. No. 103257. https://doi.org/10.1016/j.tust.2019.103257

[21] H.Y. Zhu, Z.X. Jiao, P. Zhao, X.H. Tang, S.J. Chen, and L. Tao, Physico-mechanical properties of granite after thermal treatments using different cooling media, *J. Rock Mech. Geotech. Eng.*, 2024, in press. https://doi.org/10.1016/j.jrmge.2024.10.010

[22] L. Zhu, L.G. Luo, S.H. Cui, Z.H. He, H. Wang, L.X. Zhang, and D.C. Kong, Investigation on the damage accumulation mechanisms of landslides in earthquake-prone area: role of loading-unloading cycle, *Geohaz. Mech.*, 2024, in press. https://doi.org/10.1016/j.ghm.2024.11.002

[23] Z.H. Dong, M.F. Cai, C. Ma, P.T. Wang, and P. Li, Rock damage and fracture characteristics considering the interaction between holes and joints, *Theor. Appl. Fract. Mec.*, 133(2024), art. No. 104628. https://doi.org/10.1016/j.tafmec.2024.104628

[24] R.F. Yuan and Y.H. Li, Fractal analysis on the spatial distribution of acoustic emission in the failure process of rock specimens, *Int. J. Miner. Metall. Mater.*, 16(2009), No. 1, p. 19-24. https://doi.org/10.1016/S1674-4799(09)60004-2